\documentclass[letter,
               keeplastbox,   
               ]{jacow}
\usepackage{pdfpages,multirow,ragged2e} %
\makeatletter%
	\ifboolexpr{bool{xetex}}
	 {\renewcommand{\Gin@extensions}{.pdf,%
	                    .png,.jpg,.bmp,.pict,.tif,.psd,.mac,.sga,.tga,.gif,%
	                    .eps,.ps,%
	                    }}{}
\makeatother

\ifboolexpr{bool{xetex} or bool{luatex}} 
 {}                                      
 {\usepackage[utf8]{inputenc}}           

\usepackage[USenglish]{babel}

\ifboolexpr{bool{jacowbiblatex}}%
 {%
  \addbibresource{jacow-test.bib}
  \addbibresource{biblatex-examples.bib}
 }{}
\begin{document}

\title{Beam Dynamics Studies at the PIP-II Injector Test Facility \thanks{This work was supported by the U.S. Department of Energy under contract No. DE-AC02-07CH11359}}

\author{J.-P. Carneiro\thanks{carneiro@fnal.gov}, B. Hanna, E. Podzeyev, L. Prost, A. Saini, A. Shemyakin}
	
\maketitle

\begin{abstract}
   A series of beam dynamic studies were performed in 2020-2021 at the PIP-II Injector Test Facility (PIP2IT) that has been built to validate the concept of the front-end of the PIP-II linac being constructed at Fermilab. PIP2IT is comprised of a 30-keV H- ion source, a 2 m-long Low Energy Beam Transport (LEBT), a 2.1- MeV CW RFQ, followed by a 10-m Medium Energy Beam Transport (MEBT), 2 cryomodules accelerating the beam to 16 MeV and a High-Energy Beam Transport (HEBT) bringing the beam to a dump. This paper presents beam dynamics - related measurements performed at PIP2IT such as the Twiss parameters with Allison scanners, beam envelopes along the injector, and transverse and longitudinal rms emittance reconstruction. These measurements  are compared with predictions from the beam dynamics code Tracewin.
\end{abstract}

\section{INTRODUCTION}
The PIP-II linac is an 800~MeV, 2~mA H$^{-}$ CW-capable superconducting (SC) linac for injection into the Booster\cite{pip2}. A model of the front-end of the PIP-II linac, the PIP-II Injector Test Facility (PIP2IT), has been built at Fermilab and commissioned from Summer 2020 to Spring 2021~\cite{eduard}. After a description of the PIP2IT injector, this paper presents beam dynamics measurements performed during the commissioning.         

\section{PIP2IT Overview}
\begin{figure*}[t]
    \centering
    \includegraphics*[width=\textwidth]{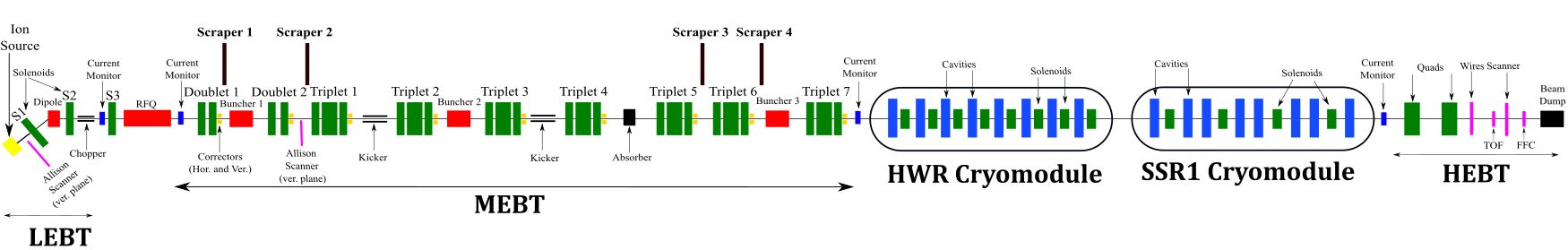}
    \caption{Layout of PIP2IT} 
    \label{fig:pip2it}
\end{figure*}

Figure~\ref{fig:pip2it} shows an overview of the PIP2IT injector. The injector is made of an ion source, a Low Energy Beam Transport (LEBT) that matches the beam into a 162.5~MHz Radiofrequency Quadrupole (RFQ), a Medium Energy Beam Transport (MEBT) that prepares the beam for injection into two SC cryomodules and a High Energy Beam Transport (HEBT) that brings the beam to a dump. The overall length of the facility is around 35 m. 

The PIP2IT ion source operates at 30~kV with long pulses (typically few ms) and at 20~Hz. The beam is focused into the RFQ with 3 solenoids and, as indicated in  Figure~\ref{fig:pip2it}, a dipole is located between the first and second solenoid to deviate the beam by a 30$^{o}$ angle. A beam chopper is located in the LEBT between the second and third solenoid and cuts pulses of up to 0.55 ms. 
As indicated in~\cite{lionel}, the LEBT operates in an original scheme. It
is kept neutralized up to the middle of the second solenoid and un-neutralized downstream of the second solenoid in the portion that contains the chopper. The RFQ accelerates the beam to an energy of 2.1~MeV. The MEBT has two main purposes: first, it performs a bunch-by-bunch selection using two kickers (that deviates the kicked beam into an absorber, as shown in Figure~\ref{fig:pip2it}) and, second, it matches the beam into the first cryomodule using 2 doublets, 5 triplets and 3 bunchers operating at 162.5~MHz. The first cryomodule contains 8  SC Half-Wave Resonators (HWR) cavities operating at 162.5~MHz, and the second cryomodule contains 8 Single-Spoke Resonators (SSR1) cavities operating at 325~MHz. The transverse focusing in the cryomodules is performed with 8 SC solenoids (HWR) and 4 SC solenoids (SSR1). The HEBT has 2 quads to transport the beam from the exit of the SSR1 cryomodule to the dump. Two correctors (horizontal and vertical) are associated with each solenoid, doublet and triplet. The PIP2IT injector has been designed to deliver a beam energy of 25~MeV at the dump and an average current of 2~mA (decreased from 5~mA by the MEBT kickers) at a maximum pulse length of 0.55~ms and 20 Hz.  

\subsection*{Diagnostics}
Five current monitors are installed along the PIP2IT injector: two in the LEBT (at the ion source exit and RFQ entrance), two in the MEBT (at the RFQ exit and at the HWR entrance) and one at the exit of the SSR1 cryomodule. The beam dump allows also for the monitoring of the current at the end of the injector. Each doublet and triplet of the MEBT and each solenoid in the cryomodules has an associated Beam Position Monitor (BPM). The HEBT contains also 3 BPMs. Two Allison scanners allow for vertical phase space measurements at the ion source exit and downstream the second MEBT doublet. In order to protect the injector from unexpected beam deviations, 4 sets of 4 scrapers (Vertical Top/Bottom, Horizontal Right/Left) are installed in the MEBT. The MEBT scrapers are also used to perform beam size measurements. Two Wire Scanners are installed in the HEBT which allow to perform beam size measurements in both horizontal and vertical planes. A movable BPM (Time-of-Flight, TOF) installed in the HEBT is used for energy measurement. A Fast Faraday Cup (FFC) located at the end of the HEBT is used for bunch length measurements. 

\subsection*{Injector Settings}

The first 3 HWR cavities were not operational during the commissioning of the injector because of a frequency offset for the first 2 cavities and a coupler issue with the third one. Furthermore, due to multipacting, the last 2 HWR cavities had to be operated at lower accelerating gradient than anticipated (respectively 8.5~MV/m and 8~MV/m vs. 9.7~MV/m). In order to compensate for that, the beam was longitudinally matched from the MEBT into the fourth HWR cavity by reaching a longitudinal waist at its entrance with a proper adjustment of the MEBT Bunchers. 

\section{Beam dynamics model}
\begin{figure}[h]
\centering
\includegraphics[width=0.9\columnwidth]
{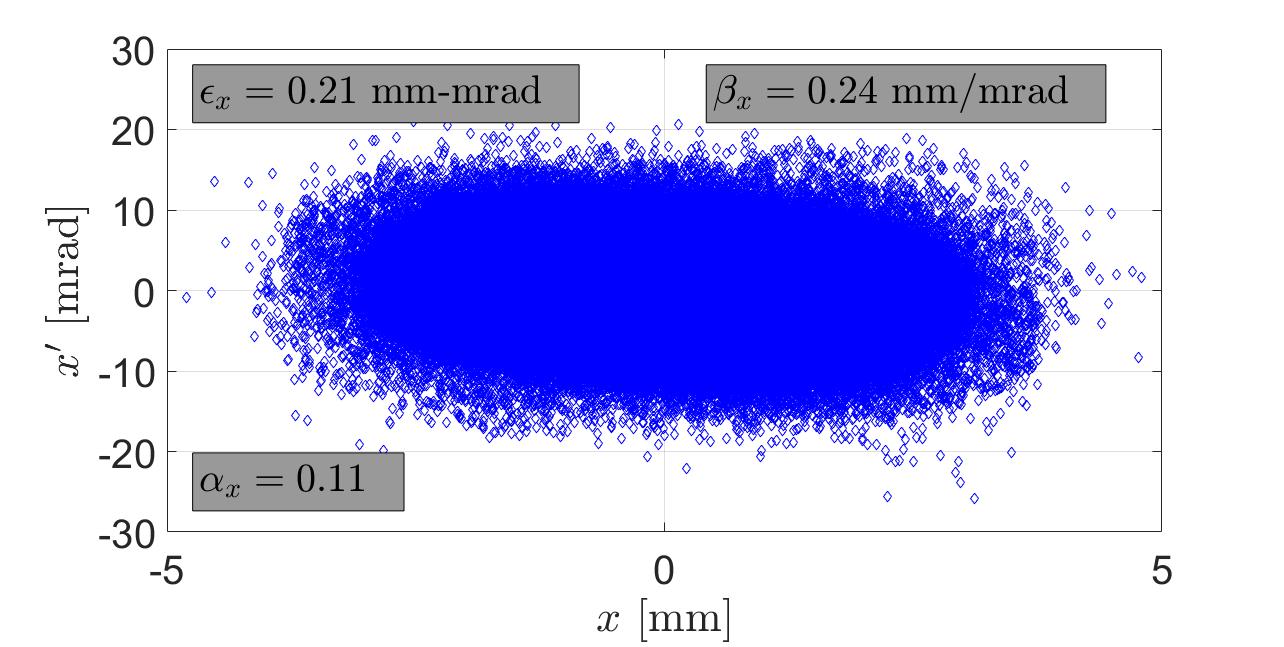} 
\put(-220,95){$\textbf{(a)}$} \quad
\includegraphics[width=0.9\columnwidth]
{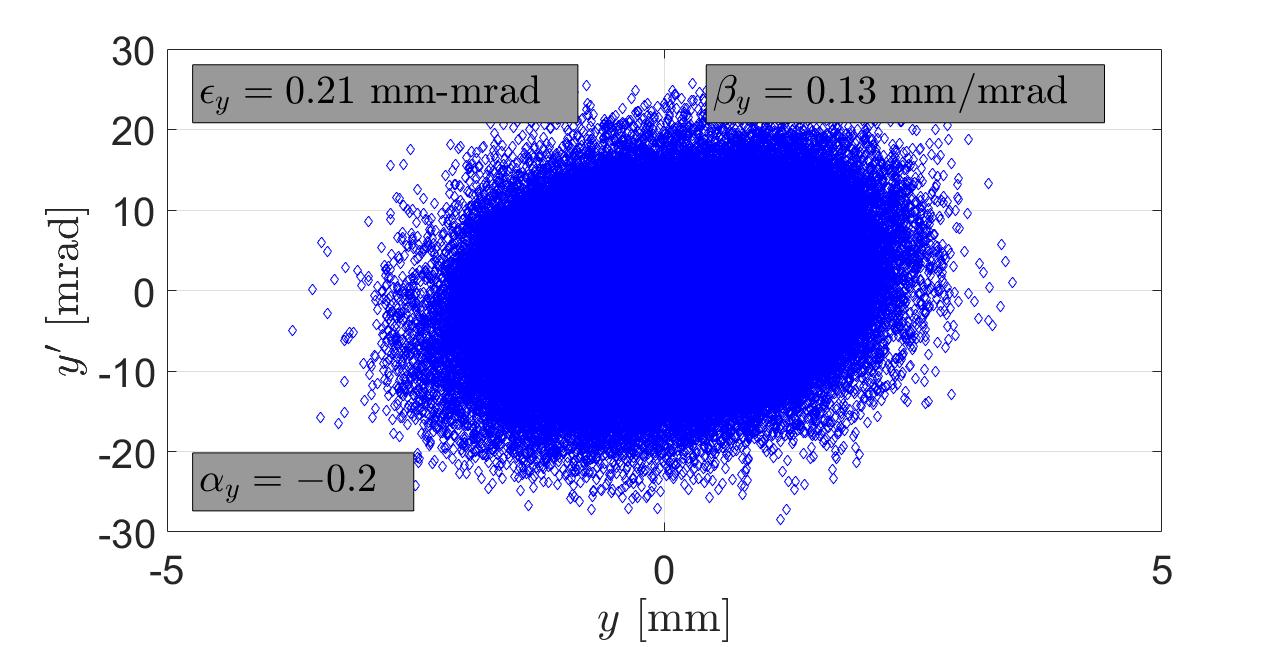} 
\put(-220,95){$\textbf{(b)}$} \quad
\includegraphics[width=0.9\columnwidth]
{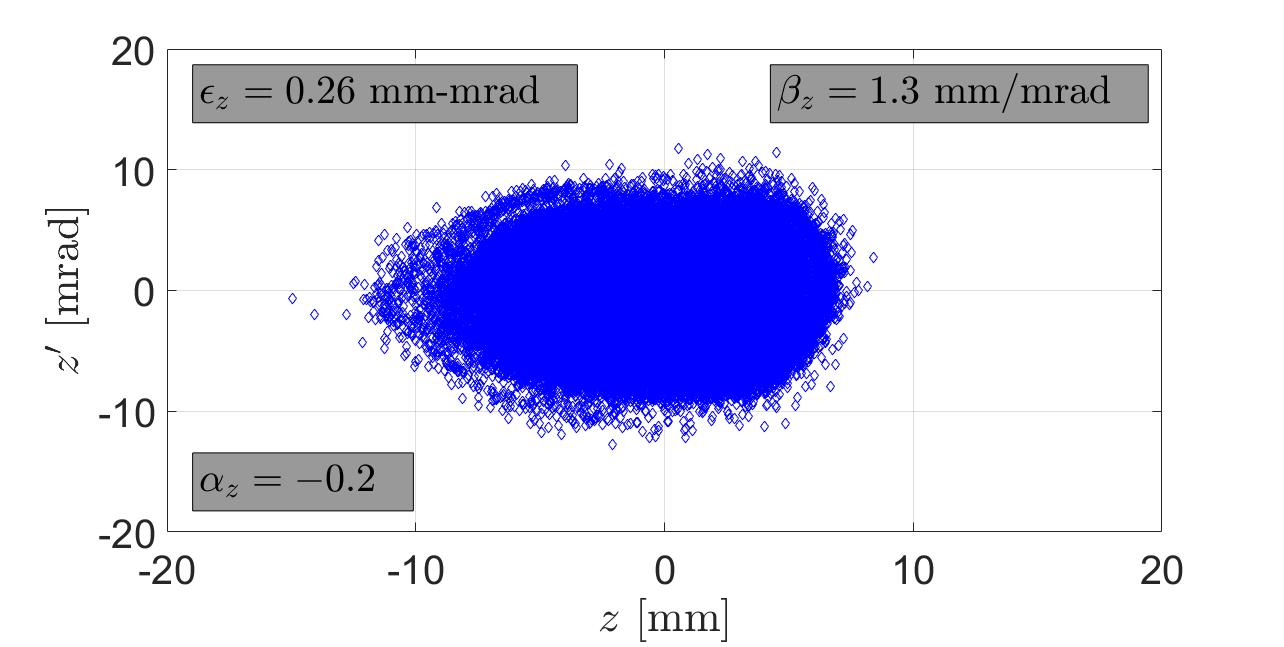}     
\put(-220,95){$\textbf{(c)}$} 
\caption{(a) Horizontal (c) Transverse (c) Longitudinal phase spaces at the RFQ exit at 5~mA with $1.1\cdot10^{6}$ macro-particles.} 
\label{fig:dist_mebt}
\end{figure}

A full 3D start-to-end model of the PIP2IT injector (from the ion source to the dump) has been build with \texttt{Tracewin}~\cite{tw}. All solenoids (LEBT, HWR and SSR1), quadrupoles (MEBT, HEBT) and correctors have been implemented in the code as 3D fields, including the LEBT dipole. All cavities (Bunchers, HWR and SSR1) have also been implemented as 3D fields. The RFQ has been modeled with~\texttt{Toutatis}~\cite{tw}. The phase spaces of the distribution at the RFQ exit is shown in Figure~\ref{fig:dist_mebt} in the 3 planes and for $1.1\cdot10^{6}$ macro-particles at 5~mA.
 
The \texttt{Tracewin} simulations start with a 4D Gaussian distribution made of $1.5\cdot10^{6}$ macro-particles cut at 6-Sigma generated by the code at the ion source exit at about 6.8~mA. The Twiss parameters used to generate the distribution are those measured with the Allison scanner located at the ion source exit for nominal operation of the injector. The beam is scrapped in the first solenoid of the LEBT and a distribution of about $1.1\cdot10^{6}$ at 5~mA reaches the end of the RFQ, as shown in Figure~\ref{fig:dist_mebt}. When the injection from the LEBT into the RFQ is optimized, the beam transmission through the RFQ is measured at about 99\%, which matches the RFQ transmission predicted by \texttt{Toutatis}.

\section{Beam Dynamics Measurement}
We present in this section MEBT beam size measurements and HEBT transverse emittance measurements performed during the commissioning for the beam energy measured with the TOF of 16~MeV (16.2~MeV expected  from \texttt{Tracewin}) and for a short pulse of 10~$\mu$s. For these measurements, the MEBT kickers were removing about 50\% of the beam and the MEBT scrapers 1 and 2 another 5\% each. We consider that about 2\% of the beam was lost in the cryomodules making an average beam current of about 2~mA reaching the HEBT. 

\subsection*{MEBT Envelope}
\begin{figure}[h]
    \centering
    \includegraphics*[width=\columnwidth]{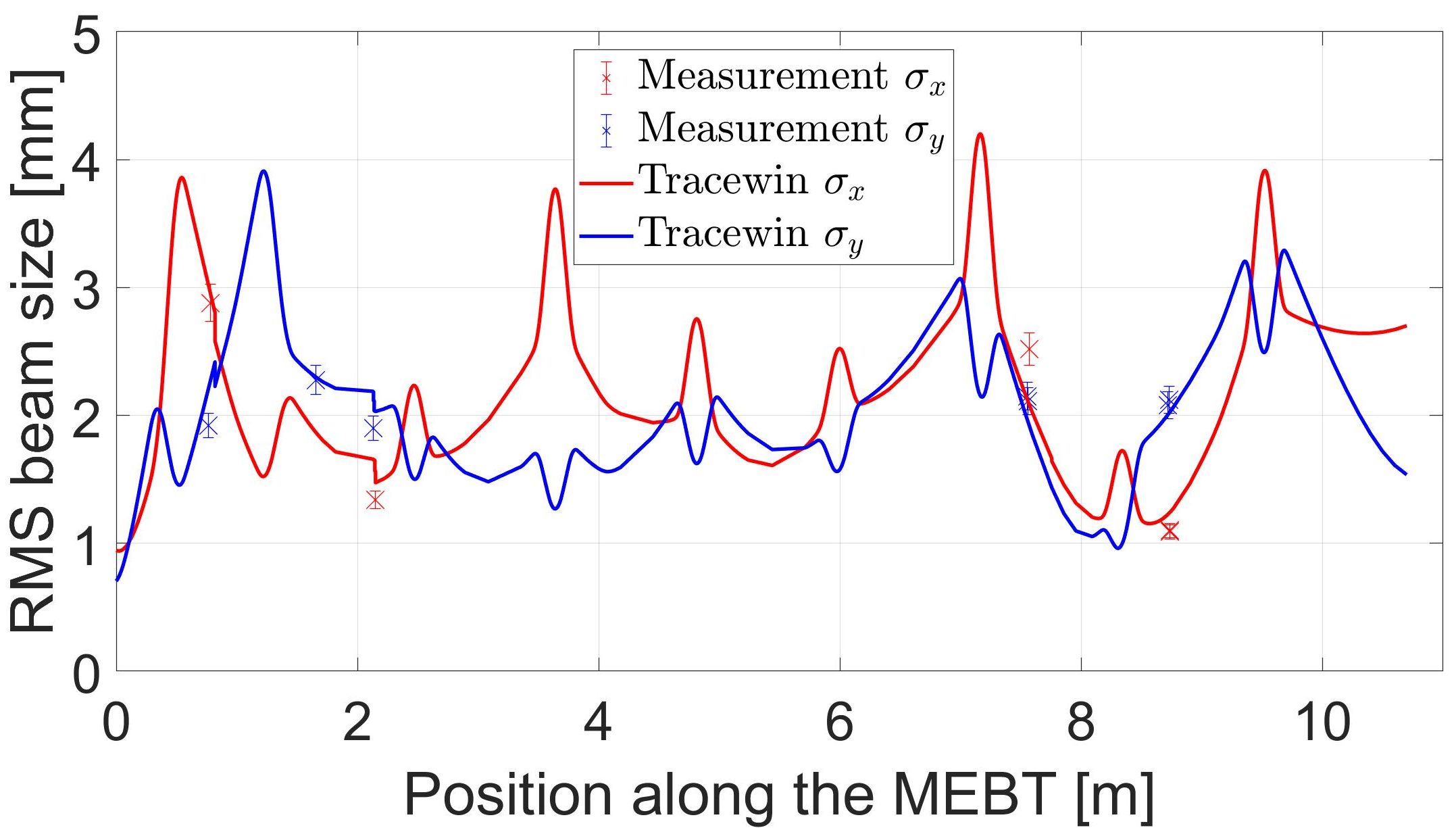}
    \caption{Beam size measurements along the MEBT and comparison with \texttt{Tracewin}.} 
    \label{fig:envelope}
\end{figure}

Figure~\ref{fig:envelope} shows the measured beam sizes along the MEBT compared with simulations from~\texttt{Tracewin}. 

The beam size measurements were performed at the 4 MEBT scrapers and at the Allison scanner. Simulations were performed with the input distribution described in Figure~\ref{fig:dist_mebt} with a transverse emittance increased to 0.25~mm-mrad matching the emittance measured in the vertical plane by the MEBT Allison scanner. At that time, we believe that the injector may not have been optimized to its best transverse emittance which usually gets down to 0.2~mm-mrad in the MEBT. A good agreement is shown in Figure~\ref{fig:envelope} between the measured beam sizes and the expected ones from \texttt{Tracewin}.

Similar beam envelope measurement in the MEBT with corresponding \texttt{Tracewin} simulations have been presented in 2018 in~\cite{arun} with the MEBT quadrupoles modeled as hard edge quadrupoles. The quadrupole strengths in 2018 were obtained from beam-based measurement and were found to be about 5\% lower than those determined by direct magnetic measurement. Furthermore, the Twiss parameters of the input MEBT distribution for the 2018 simulations were free parameters re-calculated with ~\texttt{Tracewin} to match the MEBT beam envelope measurements.

We found that using 3D quadrupole fields in \texttt{Tracewin} with the measured magnetic calibration for each individual quadrupole allows to have a good agreement with the measured MEBT beam sizes, as presented 
in Figure~\ref{fig:envelope}. It is noteworthy that the simulations presented in  Figure~\ref{fig:envelope} are, as above-mentioned,  start-to-end starting with a 4D Gaussian beam at the ion source exit. Following these results, we consider that \texttt{Tracewin} gives an accurate start-to-end model (from the ion source to the end of the MEBT) of the beam dynamics in the front-end of PIP2IT.

\subsection*{HEBT Emittances}
\begin{figure}[t]
\centering
\includegraphics[width=0.9\columnwidth]
{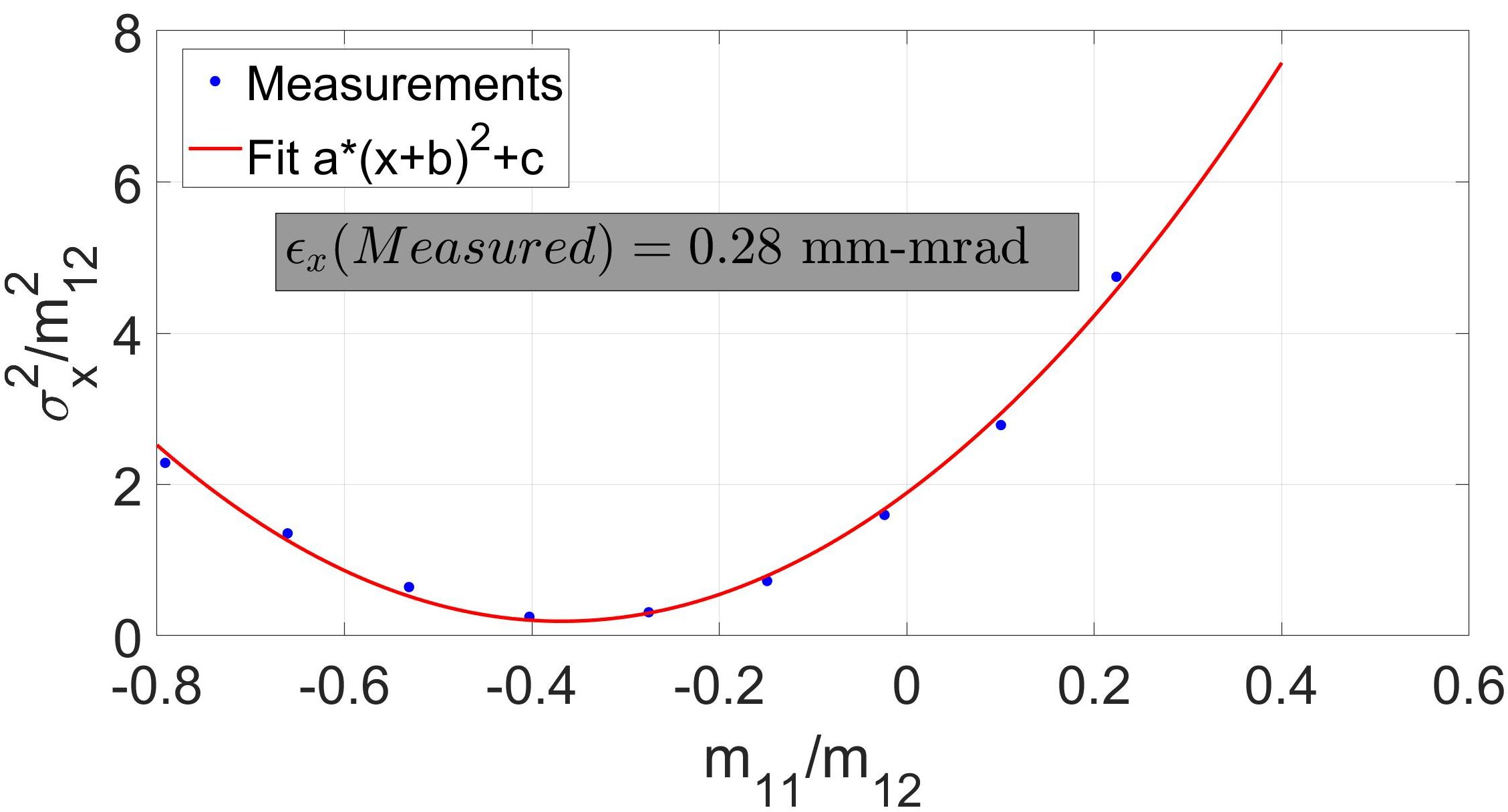} 
\put(-220,95){$\textbf{(a)}$} \quad
\includegraphics[width=0.9\columnwidth]
{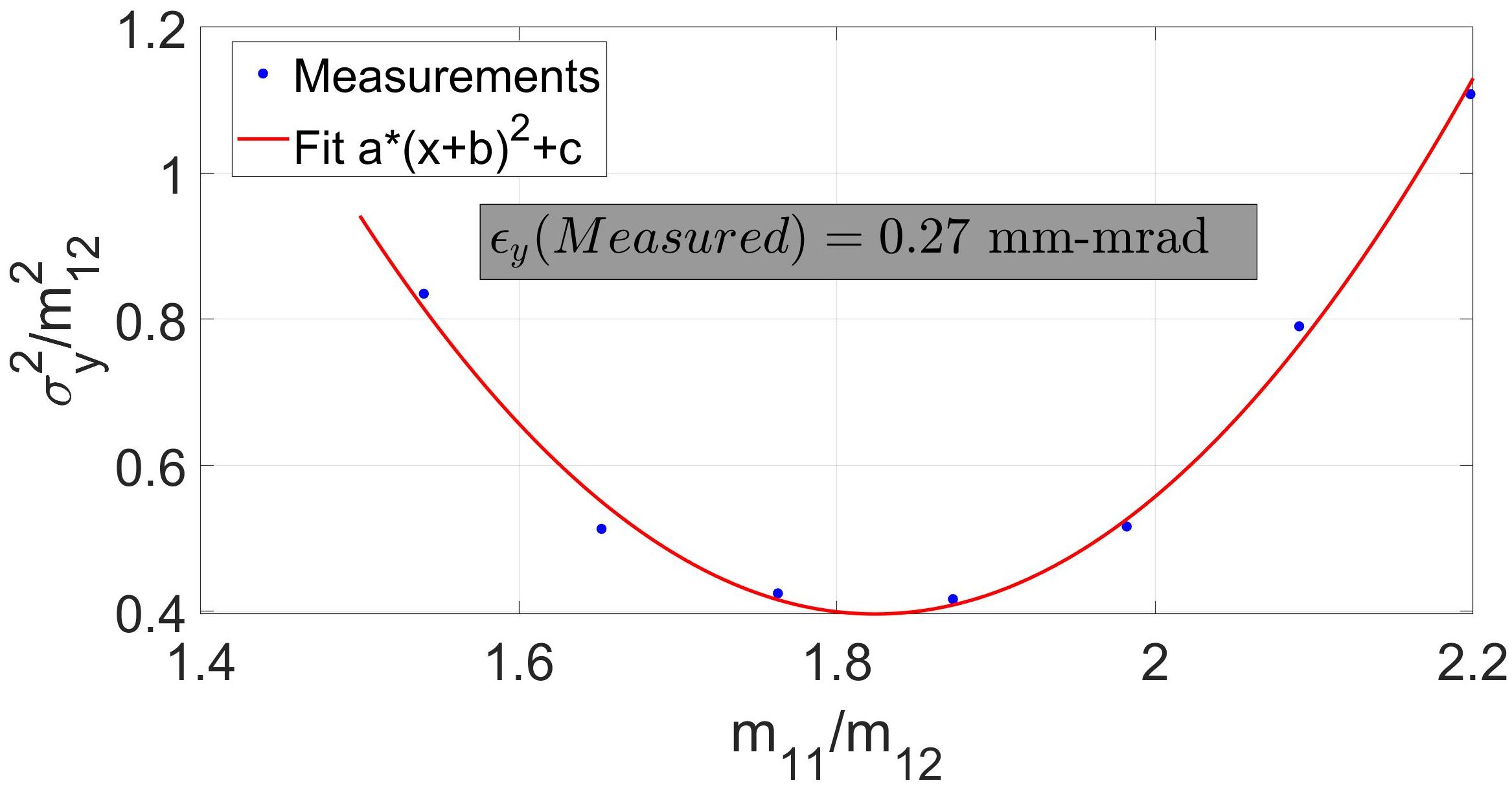} 
\put(-220,95){$\textbf{(b)}$} \quad
\caption{(a) Horizontal and (b) Vertical quadrupole scans at the HEBT Wire Scanners.} 
\label{fig:hebt_scans}
\end{figure}

Figure~\ref{fig:hebt_scans}(a) and  Figure~\ref{fig:hebt_scans}(b) show respectively the horizontal and vertical quadrupole scans performed in the HEBT using the first HEBT quad and first wire scanner (vertical emittance) and the second HEBT quad and second wire scanner (horizontal emittance). The $m_{11}$, $m_{12}$ parameters are the transfer matrix elements from the quadrupole to the wire scanner. The measured emittances in the HEBT of 0.28/0.27~mm-mrad for respectively the horizontal/vertical planes are in good agreement with the expected ones from~\texttt{Tracewin} of 0.25/0.32~mm-mrad for respectively the horizontal/vertical planes and for 85\% of the phase-spaces. A longitudinal emittance performed in the HEBT during the commissioning (using the FFC and the phase of the last SSR1 cavity) is presented in~\cite{mathias} and shows also a good agreement between the measured longitudinal emittance of 0.29~mm-mrad and expected from~\texttt{Tracewin} of 0.3~mm-mrad for 90\% of the longitudinal phase space. 

\texttt{Tracewin} does predict transverse and longitudinal beam tails in the HEBT (totaling about 10\% to 15\% of the beam). The origin of these tails are under investigation and we consider that neither the wire scanner nor the FFC are sensitive to these beam tails. Also, the reconstruction of the quad scan beam sizes using the start-to-end \texttt{Tracewin} model was not convincing, which may be due to uncertainties in the calibration of the SC solenoids.

\section{Discussion on Beam Losses}
As previously mentioned, about 2\% beam losses have been observed in the cryomodules during the operation of the PIP2IT injector. All measurements indicated that the beam losses take place mainly in the middle of the HWR cryomodule~\cite{sasha}. The start-to-end \texttt{Tracewin} model of the injector does predict 2\% beam loss in the cryomodules, but at the end of the SSR1 cryomodule. Note that \texttt{Tracewin} does not predict any  beam loss in the HWR cryomodule. The origin of discrepancy in the beam loss location is still under investigation.

\section{Conclusion}
The PIP2IT injector has been successfully commissioned from Summer 2020 to Spring 2021. Although the first 3 HWR cavities were not operational and the last 2 HWR cavities were operating at reduced gradient, a re-matching of the original beam optics allowed to transport the beam to the end of the injector, with a measured energy of about 16~MeV 
which agrees to the expected one within 1.5\%. The start (from the ion source)-to-end (up to the HEBT dump) \texttt{Tracewin} model of PIP2IT gives a good agreement with the measured beam parameters up to the end of the MEBT. Implementing 3D MEBT quad fields in Tracewin with the direct magnetic measured calibration of each quad showed a significant improvement in the accuracy of the Tracewin model in the MEBT.

\section{Acknowledgment}
We would like to thank the team members from the Accelerator and Technical Division for their help and expertise on the operation of the injector. We especially thank Vic Scarpine and the instrumentation group for their help with the diagnostics.

\ifboolexpr{bool{jacowbiblatex}}%
	{\printbibliography}%
	{%
	
	
}

\end{document}